\documentclass[conference]{IEEEtran}
\IEEEoverridecommandlockouts
\usepackage{url}
\usepackage{cite}
\usepackage{amsmath,amssymb,amsfonts}
\usepackage{algorithmic}
\usepackage{graphicx}
\usepackage{textcomp}
\usepackage{xcolor}

\usepackage{multirow}
\usepackage{stackengine}

\usepackage{hyperref}
\usepackage{footmisc}

\def\BibTeX{{\rm B\kern-.05em{\sc i\kern-.025em b}\kern-.08em
    T\kern-.1667em\lower.7ex\hbox{E}\kern-.125emX}}
\begin{document}

\title{Mixture of Mixups for Multi-label Classification \\ of Rare Anuran Sounds
\thanks{This work is supported by AI@IMT, OSO-AI, GdR IASIS, WeAMEC project PETREL, and CNRS MITI CAPTEO. 
}
}

\author{
\IEEEauthorblockN{Ilyass Moummad}
\IEEEauthorblockA{\textit{IMT Atlantique, CNRS, Lab-STICC} \\
ilyass.moummad@imt-atlantique.fr
}
\and
\IEEEauthorblockN{Nicolas Farrugia}
\IEEEauthorblockA{\textit{IMT Atlantique, CNRS, Lab-STICC} \\
nicolas.farrugia@imt-atlantique.fr
}
\and
\IEEEauthorblockN{Romain Serizel}
\IEEEauthorblockA{\textit{University of Lorraine, INRIA, LORIA} \\
romain.serizel@loria.fr
}
\and
\IEEEauthorblockN{}
\IEEEauthorblockA{\hspace{25mm}}
\and
\IEEEauthorblockN{Jeremy Froidevaux}
\IEEEauthorblockA{\textit{CREA Mont-Blanc, MNHN} \\
jeremy.froidevaux@mnhn.fr
}
\and
\IEEEauthorblockN{Vincent Lostanlen}
\IEEEauthorblockA{\textit{Ecole Centrale Nantes, CNRS, LS2N} \\
vincent.lostanlen@ls2n.fr
}
}

\maketitle

\begin{abstract}
Multi-label imbalanced classification poses a significant challenge in machine learning, particularly evident in bioacoustics where animal sounds often co-occur, and certain sounds are much less frequent than others. This paper focuses on the specific case of classifying anuran species sounds using the dataset AnuraSet, that contains both class imbalance and multi-label examples. To address these challenges, we introduce Mixture of Mixups (Mix2), a framework that leverages mixing regularization methods Mixup, Manifold Mixup, and MultiMix. Experimental results show that these methods, individually, may lead to suboptimal results; however, when applied randomly, with one selected at each training iteration, they prove effective in addressing the mentioned challenges, particularly for rare classes with few occurrences. Further analysis reveals that the model trained using Mix2 is also proficient in classifying sounds across various levels of class co-occurrences.
\footnote{Our code is released at: \url{https:/github.com/ilyassmoummad/Mix2}}
\end{abstract}

\begin{IEEEkeywords}
Bioacoustics, deep learning, mixup, multi-label classification, class imbalance.
\end{IEEEkeywords}

\section{Introduction}


Machine learning faces a significant challenge in the domain of multi-label classification, where instances can belong to multiple classes simultaneously~\cite{ml}. The issue of class imbalance, wherein certain labels are disproportionately represented is potentially adding to the complexity. This latter problem can result in overfitting of frequent classes and underfitting of rare ones\cite{iml}. Such challenges are notably prominent within the field of bioacoustics, where multi-label scenarios and class imbalance are frequently encountered~\cite{multibird, birb}.

Bioacoustics is a cross-disciplinary science that combines biology and acoustics to study sounds produced by animals. It provides insights into animal behaviour, physiological status, abundance and distribution and, at the community level, information on species diversity and interactions ~\cite{bioacoustics, app}. Birds have been extensively studied in the field of bioacoustics due to their diverse and easily detectable vocalizations in the audible range, making them a prominent and convenient model taxon for research in this field. Many bioacoustic benchmarks, tailored for machine learning applications and species detection/recognition, focus on bird sounds to address various challenges within the field, such as BirdCLEF for species classification~\cite{birdclef}, MetaAudio for few-shot classifcation~\cite{metaaudio}, DCASE 2022 and 2023 for bioacoustic event detection~\cite{task5,ines}, or BIRB~\cite{birb}, a large scale generalization benchmark. 

Avian species are pivotal subjects in bioacoustics research and passive acoustic monitoring. Yet, other vocalizing taxa hold significant importance particularly in the broader context of wildlife monitoring and conservation efforts amidst the ongoing biodiversity crisis~\cite{monitoring}. This is particularly true for Amphibians, which are one of the most endangered taxa worldwide, with almost half at the risk of extinction~\cite{decline, anuraset}. Amphibians are highly sensitive to both land-use and climate changes, further compounded by the devastating effects of the pathogenic fungal disease chytridiomycosis~\cite{amphibian, importance}. Standardized, wide-scale monitoring is crucial to facilitate the formulation and implementation of targeted conservation strategies. In this regard, the use of acoustic tools offer promising prospects for anuran monitoring and conservation programmes. However, automatic identification of anuran species presents a significant challenge due to their nocturnal, elusive behavior, diverse habitats, seasonal influences, species diversity, limited vocalization windows, and often remote locations. While a few  studies leverage machine learning for the classification of anuran sounds~\cite{strout, alabi, akbal, colonna}, most of them rely on privately collected recordings, posing challenges for comparisons between different works.

The recent introduction of the AnuraSet dataset~\cite{anuraset} addresses this issue by providing an extensive compilation of Neotropical anuran vocalizations. This dataset currently stands as the  largest open-source dataset in anuran sound research. Encompassing two Brazilian biomes, it features a diverse collection of 42 anuran species, demonstrating strong co-occurrence, including certain underrepresented species. In addition, certain rare classes exhibit no overlap between the training and test sets. AnuraSet presents inherent challenges characterized by imbalanced distributions, multi-label instances, and noisy conditions in the natural environment. It provides a significant benchmark for addressing these challenges in the application of anuran sound classification.

In order to address such difficult generalization problems in deep learning, promising solutions are mixing regularization methods, such as Mixup~\cite{Mixup}, that dynamically generate synthetic examples during training. Mixup penalizes the loss accordingly to prevent the model from overfitting the training data and to smooth decision boundaries from class to class. This leads to better prediction outside the training examples. These methods regularize neural networks by promoting linear behavior between training samples in either input space or latent space~\cite{Mixup, manMixup, multimix}. 
In order to adress class imbalance, Remix~\cite{remix} relaxes the mixing factor by pushing the decision boundaries towards majority classes. Regarding multi-label classification, FlowMixup~\cite{flowmixup} decouples the extracted features by adding constraints to the hidden states of the models. 

In this work, we investigate three mixing regularization methods: Mixup~\cite{Mixup}, Manifold Mixup~\cite{manMixup}, and MultiMix~\cite{multimix}, to address multi-label imbalanced classification in AnuraSet dataset. Our findings show that alternating between these methods during training leads to significant improvement in classification performance, especially for rare classes.

\section{Method}

\begin{figure*}
\begin{minipage}[b]{1.0\linewidth}
  \centering
  \centerline{\includegraphics[width=1.\textwidth]{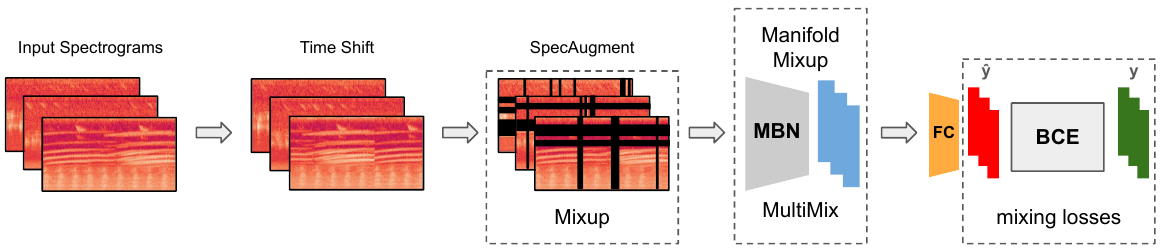}}
\end{minipage}
\caption{Overview of our system: Mix2. MBN, FC and BCE stand for MobileNetV3-Large, Fully Connected, and Binary Cross-Entropy, respectively.}
\label{fig:mix2}
\end{figure*}

In Figure~\ref{fig:mix2}, we present a comprehensive overview of our proposed system, Mix2. Subsequently, we elaborate on the distinct components of Mix2.

\subsection{Mixup}

Mixup~\cite{Mixup} is a technique that involves training on convex combinations of pairs of examples and their corresponding labels. This method induces decision boundaries that linearly transition from one class to another, resulting in a smoother estimation of uncertainty.
Although its effectiveness is not fully understood, Mixup is widely employed as a regularization technique in deep learning for supervised classification problems. 
Consider two randomly sampled instances from the dataset $\boldsymbol{x}_i$ and $\boldsymbol{x}_j$ with their corresponding labels $\boldsymbol{y}_i$ and $\boldsymbol{y}_j$.
A synthetic example $\boldsymbol{\tilde{x}}$ and its label $\boldsymbol{\tilde{y}}$ are generated via the formula:

\[ \boldsymbol{\tilde{x}} = \lambda \boldsymbol{x}_i + (1 - \lambda) \boldsymbol{x}_j \]
\[ \boldsymbol{\tilde{y}} = \lambda \boldsymbol{y}_i + (1 - \lambda) \boldsymbol{y}_j \]

where $\mathbf{\lambda}$ represents a random interpolation coefficient sampled from a Beta distribution.

\subsection{Manifold Mixup}

Manifold Mixup~\cite{manMixup} extends the concept of Mixup to the embeddings of a neural network, introducing interpolation at this level to smooth decision boundaries across multiple levels of representation. The synthetic embedding, $\boldsymbol{\tilde{h}}$, is computed using the formulas:

\[ \boldsymbol{\tilde{h}} = \lambda \boldsymbol{h}_i + (1 - \lambda) \boldsymbol{h}_j \]
\[ \boldsymbol{\tilde{y}} = \lambda \boldsymbol{y}_i + (1 - \lambda) \boldsymbol{y}_j \]

Here, $\boldsymbol{\tilde{h}}_i$ and $\boldsymbol{\tilde{h}}_j$ represent the embeddings of two instances, $\boldsymbol{y}_i$ and $\boldsymbol{y}_j$ are their corresponding labels.

\subsection{MultiMix}

Both Mixup and Manifold Mixup involve interpolating two training samples and their corresponding labels. In contrast, MultiMix~\cite{multimix} extends this concept to the entire mini-batch in the embedding space. The synthetic batch of embeddings, $\tilde{H}$, is obtained through the formulas:

\[ \mathbf{\tilde{H}} = \mathbf{\Lambda H} \]
\[ \mathbf{\tilde{Y}} = \mathbf{\Lambda Y} \]

Here, $\mathbf{H}$ is the matrix of embeddings for the original batch, $\mathbf{\Lambda}$ is a matrix of interpolation coefficients sampled from a Dirichlet distribution, $\mathbf{Y}$ is the matrix of corresponding one-hot encoded labels for the original batch, and $\mathbf{\tilde{Y}}$ is the synthetic batch of labels. The sampling occurs on the entire convex hull of the mini-batch, rather than linear segments between pairs of examples. The matrix product $\mathbf{\Lambda H}$ performs the interpolation between all samples in the batch.
Likewise, $\mathbf{\Lambda Y}$ interpolates between the corresponding labels.

For all of these methods, in practice, the mixing of targets is performed at the loss level.

\subsection{Novel contribution: Mixture of Mixups (Mix2)}
We now present Mix2, a probabilistic mixture of the three aforementioned methods.
At each SGD training iteration, Mix2 employs random selection from three mixing regularization options: Mixup, Manifold Mixup, or Multimix.
This dynamic approach ensures that the model learns from a diverse range of augmented samples, promoting robustness and generalization.

\section{Evaluation}

\subsection{Dataset}

\begin{figure}
\begin{minipage}[b]{1.0\linewidth}
  \centering
  \centerline{\includegraphics[width=1.\textwidth]{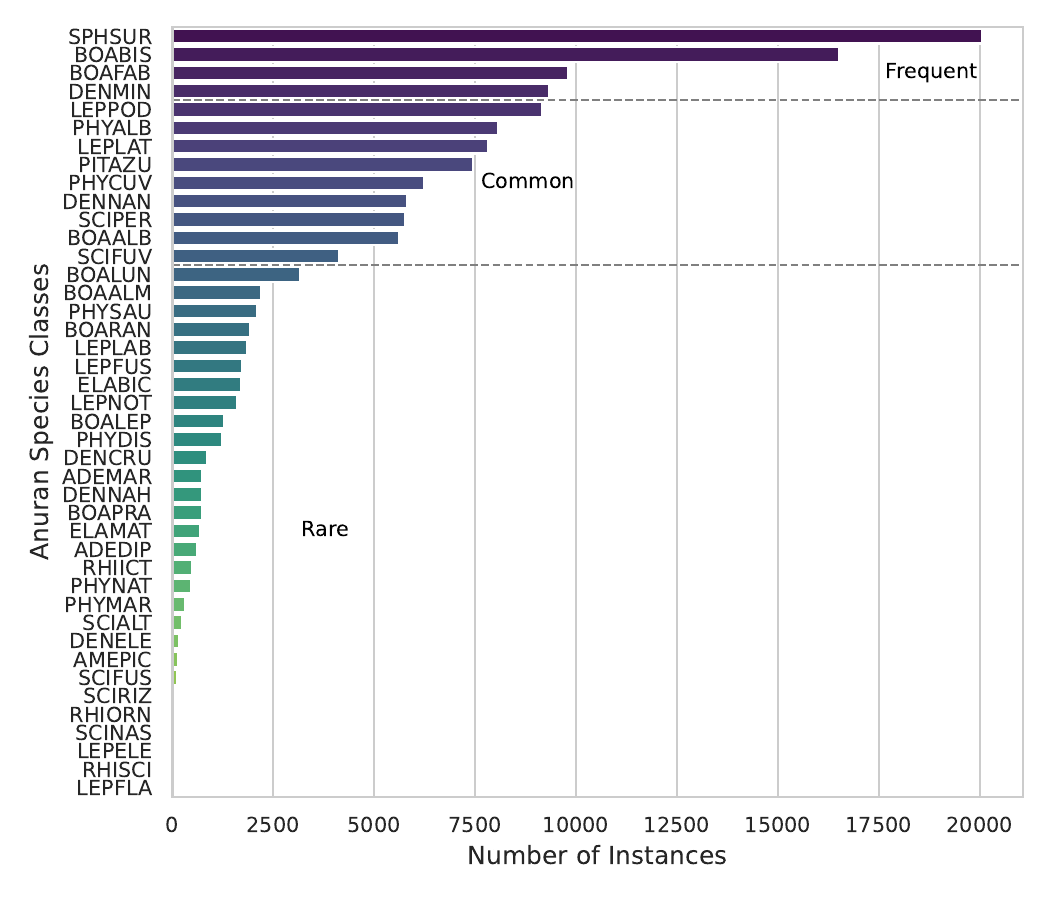}}
\end{minipage}
\caption{Histogram of class imbalance in AnuraSet.
Dashed lines denote ``frequent'' species (more than 10k instances), ``common'' species (5k--10k instances), and ``rare'' species (fewer than 5k instances).
}
\label{fig:imbalance}
\end{figure}

\begin{figure}
\begin{minipage}[b]{1.0\linewidth}
  \centering
  \centerline{\includegraphics[width=1.\textwidth]{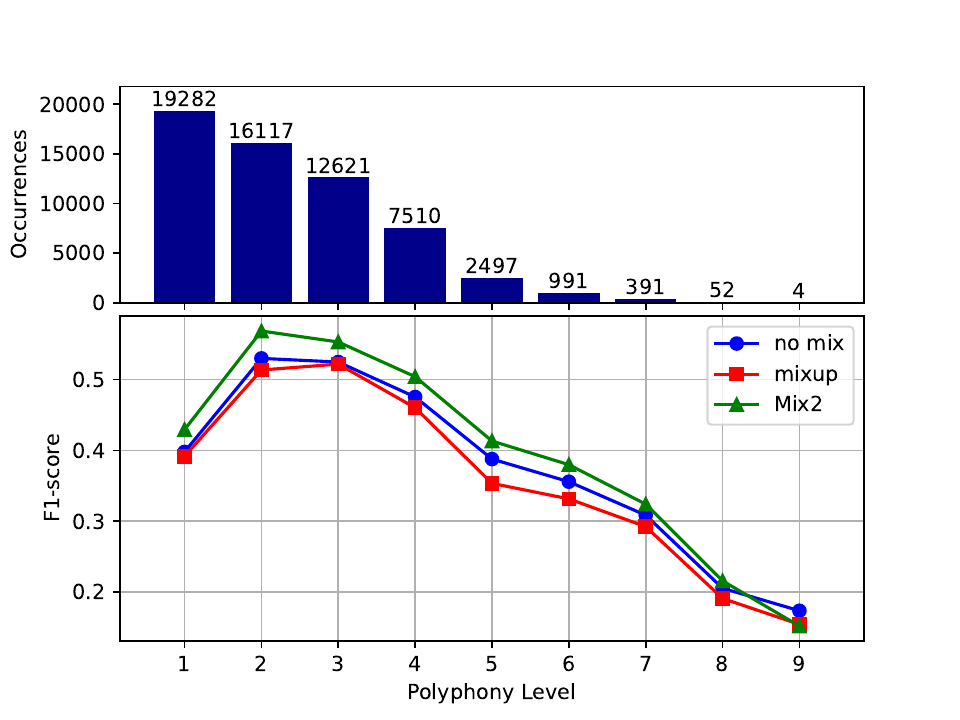}}
\end{minipage}
\caption{In AnuraSet, Mixup leads to a performance decline across various polyphony levels compared to the absence of mixing. Conversely, Mix2 demonstrates an improvement in performance across different polyphony levels.}
\label{fig:polyperf}
\end{figure}

AnuraSet~\cite{anuraset} constitutes a multi-label classification dataset comprising 93,378 3-second segments capturing anuran calls from 42 distinct species. Derived from 1,612 annotated 1-minute audio recordings, equivalent to 26.87 hours of acoustic documentation, the dataset is generated using fixed-length 3-second windows with a 1-second sliding window (75\% overlap between samples). To prevent data leakage between training and testing, the dataset is partitioned into a 2/3 train set and a 1/3 test set at the audio recording level. The upper part of Figure~\ref{fig:polyperf} illustrates the polyphony levels within the dataset, while Figure~\ref{fig:imbalance} categorizes the species into three groups: frequent (more than 10k), common (between 5k and 10k), and rare (fewer than 5k).

Given the absence of a dedicated validation set, designing one from the training set is challenging due to the overlap between examples. The split should be conducted at the audio recording level, from which examples were extracted using a sliding window. Additionally, due to the long-tail distribution, splitting classes with few occurrences into distinct subsets is difficult; this problem is already evident between the train and test sets. Consequently, creating a proper validation set poses a challenge. In line with standard machine learning benchmarks, we conduct model validation on the test set.

\subsection{Metric}

The benchmark employs the macro $F$-score, assigning equal importance to all species. It is computed as the mean of the $F$-score for each class.

\subsection{Model}

MobileNetV3, or MBN for short, is a well-established series of convolutional neural network architecture for embedded computing applications~\cite{mbn}.
The architectural refinements of MobileNetV3 include: squeeze-and-excitation blocks, residual connections, depthwise convolutions, and atrous spatial pyramid pooling.
In this paper, we propose to train a MobileNetV3-Large, which has three million parameters.


\subsection{Time--frequency representation}
We resample AnuraSet to 16 kHz. Following the AnuraSet baseline, we compute a mel-frequency spectrogram with 128 filters, a window size of 512, and a hop size of 128.

\subsection{Data augmentation}
Following the AnuraSet baseline, we apply SpecAugment~\cite{specaugment}, a data augmentation technique which sets time--frequency magnitudes to zero over multiple random rectangular regions of the mel-frequency spectrogram.
In addition, we apply random circular time shift following a previous study on machine learning for birdsong classification~\cite{ssl4birds}.

\subsection{Training}

For all our experiments, we train MobileNetV3-Large~\cite{mbn} from scratch using AdamW optimizer with a batch size of 128, a learning rate of $10^{-2}$, and a weight decay of $10^{-6}$ for 100 epochs. 
We train all our systems for 100 epochs, with the exception of the first row (10 epochs) in Table~\ref{Tab:res} for a fair comparison with the baseline systems of AnuraSet, which are trained for 10 epochs.
When we train the model using Mix2, we select Mixup, Manifold Mixup, or Multimix, with probabilities of 25\%, 50\%, and 25\%, respectively. Manifold Mixup is selected more frequently as it was the best-performing single regularization method.

\subsection{Tasks}

Certain classes exhibit no overlap between the training and test sets, with 4 found only in the training set and 2 only in the test set. Moreover, a substantial portion of the dataset lacks any active class (33,913 out of 93,378). In evaluating the benchmark, the assessment is exclusively conducted on the active classes. To gain insights, we create two subsets of the AnuraSet. The first subset involves removing the non-overlapping classes between training and testing, referred to as AnuraSet-36N (N stands for negative as it also contain examples with no class activated), with the same number of examples as AnuraSet but containing 36 classes instead of 42. The second subset, extends beyond the non-overlapping classes by also eliminating examples with silence, referred to as AnuraSet-36 (with 59,465 examples and 36 classes). In Table~\ref{Tab:subset}, we compare the performance of Mix2 with Mixup and no mixing on these two subsets.

\section{Results and Discussions}

We conduct a comprehensive ablation study to demonstrate the effect of each individual component of our proposed method that contributes to the improvement in $F$-score.

\begin{table}
\caption{Macro $F$-scores (in \%) on AnuraSet.\\We report means and standard deviations over three runs. Frequent, common, and rare classes are defined in Figure \ref{fig:imbalance}.}
\label{Tab:res}
\centering
{\begin{tabular}{lcccc}
\hline
Method & Frequent & Common & Rare & All \\
\hline
\multicolumn{5}{c}{State of the art: 11M--116M parameters, ImageNet pre-training~\cite{anuraset}} \\
ResNet18 & 61.6 & 52.0 & 14.8 & 34.9 \\
ResNet50 & 62.3 & 53.9 & 9.9 & 33.2 \\
ResNet152 & 68.4 & 56.8 & 15.7 & 37.8 \\ \\
\hline
\multicolumn{5}{c}{Ours: 3M parameters, no pre-training} \\
MobileNetV3-Large (10 epochs) & \stackunder{86.3}{\scriptsize$\pm$0.2} & \stackunder{72.4}{\scriptsize$\pm$0.3} & \stackunder{35.4}{\scriptsize$\pm$1.0} & \stackunder{48.2}{\scriptsize$\pm$0.6} \\
+90 epochs & \stackunder{88.9}{\scriptsize$\pm$0.2} & \stackunder{75.5}{\scriptsize$\pm$0.3} & \stackunder{45.6}{\scriptsize$\pm$0.6} & \stackunder{56.2}{\scriptsize$\pm$0.5} \\
+SpecAug & \stackunder{86.7}{\scriptsize$\pm$0.4} & \stackunder{74.41}{\scriptsize$\pm$0.4} & \stackunder{47.0}{\scriptsize$\pm$1.0} & \stackunder{56.6}{\scriptsize$\pm$0.7} \\
+TimeShift & \stackunder{87.0}{\scriptsize$\pm$0.2} & \stackunder{74.2}{\scriptsize$\pm$0.0} & \stackunder{46.9}{\scriptsize$\pm$1.0} & \stackunder{56.6}{\scriptsize$\pm$0.7} \\
\multicolumn{5}{l}{+Regularization} \\
\multicolumn{5}{l}{(1 method)} \\
~~~~Mixup & \stackunder{87.4}{\scriptsize$\pm$0.2} & \stackunder{74.3}{\scriptsize$\pm$0.5} & \stackunder{44.9}{\scriptsize$\pm$0.4} & \stackunder{55.2}{\scriptsize$\pm$0.4} \\
~~~~Manifold Mixup & \stackunder{86.8}{\scriptsize$\pm$0.4} & \stackunder{74.3}{\scriptsize$\pm$0.4} & \stackunder{47.0}{\scriptsize$\pm$0.8} & \stackunder{56.6}{\scriptsize$\pm$0.5} \\
~~~~MultiMix & \stackunder{88.0}{\scriptsize$\pm$0.4} & \stackunder{75.1}{\scriptsize$\pm$0.1} & \stackunder{46.1}{\scriptsize$\pm$0.2} & \stackunder{56.3}{\scriptsize$\pm$0.2} \\
\multicolumn{5}{l}{(2 methods)} \\
~~~~Mixup + Manifold Mixup & \stackunder{88.6}{\scriptsize$\pm$0.1} & \stackunder{76.6}{\scriptsize$\pm$0.2} & \stackunder{51.7}{\scriptsize$\pm$1.1} & \stackunder{60.5}{\scriptsize$\pm$0.8} \\
~~~~Mixup + MultiMix & \stackunder{88.5}{\scriptsize$\pm$1.1} & \stackunder{75.7}{\scriptsize$\pm$0.9} & \stackunder{47.8}{\scriptsize$\pm$2.1} & \stackunder{57.7}{\scriptsize$\pm$1.7} \\
~~~~Manifold Mixup + MultiMix & \stackunder{88.1}{\scriptsize$\pm$0.4} & \stackunder{75.7}{\scriptsize$\pm$0.1} & \stackunder{48.4}{\scriptsize$\pm$2.3} & \stackunder{58.0}{\scriptsize$\pm$1.7} \\
\multicolumn{5}{l}{(3 methods)} \\
~~~~Mixture of Mixups (Mix2) & \textbf{\stackunder{89.2}{\scriptsize$\pm$0.1}} & \textbf{\stackunder{77.1}{\scriptsize$\pm$0.1}} & \textbf{\stackunder{51.8}{\scriptsize$\pm$0.4}} & \textbf{\stackunder{60.8}{\scriptsize$\pm$0.3}} \\ \\
\hline
\end{tabular}
}
\end{table}      

\begin{table}
\caption{Macro $F$-scores (in \%) on AnuraSet-36N and AnuraSet-36. \\ We report means and standard deviations over three runs. \\
Frequent, common, and rare classes are defined in Figure \ref{fig:imbalance}.}
\label{Tab:subset}
\centering
\begin{tabular}{lcccc}
\hline
Method & Frequent & Common & Rare & All \\
\hline
\multicolumn{5}{c}{AnuraSet-36N} \\
no mixing & \stackunder{87.4}{\scriptsize$\pm$0.3} & \stackunder{83.9}{\scriptsize$\pm$0.2} & \stackunder{56.3}{\scriptsize$\pm$0.5} & \stackunder{65.9}{\scriptsize$\pm$0.3} \\
Mixup & \stackunder{87.2}{\scriptsize$\pm$0.4} & \stackunder{83.5}{\scriptsize$\pm$0.3} & \stackunder{55.0}{\scriptsize$\pm$0.7} & \stackunder{64.9}{\scriptsize$\pm$0.5} \\
Mix2 & \stackunder{\textbf{88.9}}{\scriptsize$\pm$\textbf{0.2}} & \stackunder{\textbf{86.6}}{\scriptsize$\pm$\textbf{0.3}} & \stackunder{\textbf{62.0}}{\scriptsize$\pm$\textbf{0.5}} & \stackunder{\textbf{70.4}}{\scriptsize$\pm$\textbf{0.3}} \\ \\
\hline
\multicolumn{5}{c}{AnuraSet-36} \\
no mixing & \stackunder{88.8}{\scriptsize$\pm$0.3} & \stackunder{84.4}{\scriptsize$\pm$0.2} & \stackunder{60.6}{\scriptsize$\pm$0.7} & \stackunder{69.0}{\scriptsize$\pm$0.4} \\
Mixup & \stackunder{88.5}{\scriptsize$\pm$0.3} & \stackunder{83.9}{\scriptsize$\pm$0.2} & \stackunder{58.8}{\scriptsize$\pm$0.7} & \stackunder{67.7}{\scriptsize$\pm$0.4} \\
Mix2 & \textbf{\stackunder{89.5}{\scriptsize$\pm$0.1}} & \textbf{\stackunder{86.8}{\scriptsize$\pm$0.2}} & \textbf{\stackunder{67.4}{\scriptsize$\pm$0.6}} & \textbf{\stackunder{74.2}{\scriptsize$\pm$0.5}} \\ \\
\hline
\end{tabular}
\end{table}

The upper part of Table~\ref{Tab:res} shows the three proposed baselines from the AnuraSet paper~\cite{anuraset}. These baselines involve fine-tuning three residual convolutional architectures ResNet18, ResNet50, ResNet152 pre-trained on ImageNet~\cite{imagenet}, using the training set of AnuraSet sampled at a rate of 22,050 Hz. The fine-tuning is conducted for 10 epochs, using SpecAugment with a masking strategy covering up to 60 time bands and 8 frequency bands.
The second part of the table displays our results, where each row illustrates the impact of adding a specific technique on top of all the previous ones. Training longer (100 vs 10 epochs) contributes to higher performance. The addition of SpecAugment provides a slight boost in scores, particularly for rare classes. Introducing a single mixing regularization negatively affects performance. Conversely, combining two methods, especially Mixup and Manifold Mixup, results in a substantial performance boost. Similarly, Mix2, incorporating all three methods, achieves the highest score with more than 4\% gain in macro $F$-score. The gain in $F$-score is also evident in the rare group, highlighting the efficiency of our proposed method to handle class imbalance in multi-label classification.

This phenomenon is intriguing, given that the individual regularization methods, when applied alone, tend to decrease performance (with the exception of Manifold Mixup). We speculate that combining different mixing regularizations with a certain probability leads to a more diverse set of augmented training examples, capturing a broader range of data variations and enhancing the model's robustness.

In addition, we experimented with weighted sampling by oversampling rare combinations of classes. However, this approach yielded worse results than the baseline and was therefore excluded from the study.

In Figure~\ref{fig:polyperf}, we study the macro $F$-score score of examples with different polyphony levels ranging from 1 to 9 for Mix2, Mixup, and when no mixing is applied. We observe a similar trend where Mixup performs worse than when it is not applied, and Mix2 noticeably improves the performance, a finding worthy of future investigation. Error bars are shown at each point, with the standard deviation being very small across runs. 

Table~\ref{Tab:subset} compares the performance of Mix2 with Mixup and no mixing the impact of removing the non-overlapping classes from the dataset (AnuraSet-36N) in the first part, as well as when also the segments with no class activated are removed (AnuraSet-36) in the second part. Similar to Table~\ref{Tab:res}, this result confirms that Mix2 effectively addresses the challenges associated with multi-label classification and class imbalance, rather than uncovering any hidden issues within the inherently ambiguous classification problem of AnuraSet (examples with no class and non-overlapping train/test classes).

\section{Conclusions and Perspectives}

This study addresses the challenges of multi-label imbalanced classification, particularly within the domain of bioacoustics focusing on anuran species sound classification using the AnuraSet dataset. The proposed Mix2 system, combining mixing regularization methods Mixup, Manifold Mixup, and MultiMix, to capture a wide range of data variation, has demonstrated significant efficacy in addressing imbalance and multi-label. 

For future work, we want to investigate whether combining different mixing strategies learn better representations in self-supervised learning, and whether the learned representations generalize well in out-of-distribution settings. Furthermore, we aim to integrate ecological information on diel temporal activity patterns of different species and account for potential temporal acoustic niche segregation \cite{krause} to only mix sounds of species co-occuring spatially and temporally.

The presence of non-overlapping rare classes in the train/test sets poses a challenge. One promising avenue for addressing this challenge is the exploration of few-shot learning, where the model is trained with limited annotations for these rare classes. Additionally, zero-shot learning, leveraging multi-modal approaches such as language, provides another interesting direction for research in mitigating the impact of non-overlapping rare classes. Another direction is open-set classification, where the objective is to classify events from unknown classes~\cite{openset}.



\bibliographystyle{IEEEbib}
\bibliography{refs}

\end{document}